\documentclass[12pt]{iopart}

\usepackage{graphicx}
\usepackage{caption}
\usepackage{subcaption}
\usepackage{booktabs}
\usepackage[pdftex, hidelinks]{hyperref}
\usepackage{placeins}

\expandafter\let\csname equation*\endcsname\relax
\expandafter\let\csname endequation*\endcsname\relax
\usepackage{amsmath}

\usepackage{url}
\RequirePackage{doi}
\usepackage[
backend=biber,
citestyle=numeric-comp,
bibstyle=ieee,
sorting=none,
minbibnames=3,
maxbibnames=3,
giveninits=true,
url=false,
]{biblatex}
\usepackage{microtype}
\begin{document}

\title[Deep Learning Models for the AMS Electromagnetic Calorimeter]{A Comparison of Deep Learning Models for Proton Background Rejection with the AMS Electromagnetic Calorimeter}

\author{R. K. Hashmani$^1$\footnote{Present address:
Department of Physics, University of Wisconsin–Madison, Madison, WI 53706, USA.}, E. Akbaş$^1$ and M. B. Demirköz$^2$}

\address{$^1$ Department of Computer Engineering, 
Middle East Technical University (METU), Ankara 06800, Turkey}
\address{$^2$ Department of Physics, Middle East Technical University (METU), Ankara 06800, Turkey}
\ead{raheem.hashmani@cern.ch}

\begin{abstract}
The Alpha Magnetic Spectrometer (AMS) is a high-precision particle detector onboard the International Space Station containing six different subdetectors. The Transition Radiation Detector and Electromagnetic Calorimeter (ECAL) are used to separate electrons/positrons from the abundant cosmic-ray proton background.

The positron flux measured in space by AMS falls with a power law which unexpectedly softens above 25 GeV and then hardens above 280 GeV. Several theoretical models try to explain these phenomena, and a purer measurement of positrons at higher energies is needed to help test them. The currently used methods to reject the proton background at high energies involve extrapolating shower features from the ECAL to use as inputs for boosted decision tree and likelihood classifiers. We present a new approach for particle identification with the AMS ECAL using deep learning (DL). By taking the energy deposition within all the ECAL cells as an input and treating them as pixels in an image-like format, we train an MLP, a CNN, and multiple ResNets and Convolutional vision Transformers (CvTs) as shower classifiers.

Proton rejection performance is evaluated using Monte Carlo (MC) events and ISS data separately. For MC, using events with a reconstructed energy between 0.2 – 2 TeV, at 90\% electron accuracy, the proton rejection power of our CvT model is more than 5 times that of the other DL models. Similarly, for ISS data with a reconstructed energy between 50 – 70 GeV, the proton rejection power of our CvT model is more than 2.5 times that of the other DL models.
\end{abstract}

\vspace{2pc}
\noindent{\it Keywords}: Alpha Magnetic Spectrometer, Electromagnetic Calorimeter, Astroparticle Physics, Deep Learning, Vision Transformers
%
%
\maketitle

\section{Introduction}

The Alpha Magnetic Spectrometer (AMS) is a high-precision particle physics detector onboard the International Space Station (ISS). Installed on 19 May 2011, its main objectives include searching for antimatter, investigating dark matter, and analyzing cosmic rays. To achieve this, the AMS is equipped with 6 detectors: the gaseous Xe/CO\textsubscript{2} Transition Radiation Detector (TRD), two Time of Flight (ToF) Counters, the nine layer Silicon Tracker (Tracker), the aerogel/NaF Ring Imaging Cherenkov (RICH), and the lead/scintillator fiber Electromagnetic Calorimeter (ECAL). At the center, a permanent Nd-Fe-B magnet is used to generate a magnetic field of 0.15 T \cite{Ting2013TheStation}. A state-of-the-art monitoring system is used to ensure the detectors and electronics are always working in optimal conditions \cite{Hashmani2023NewExperiment}. AMS has been continuously collecting cosmic ray particles since its inauguration and has collected over 231 billion cosmic ray events \cite{AMSCollaboration2023AMS-02}.

\begin{figure}[htb]
     \centering
     \begin{subfigure}[b]{0.49\textwidth}
         \centering
         \includegraphics[width=\textwidth]{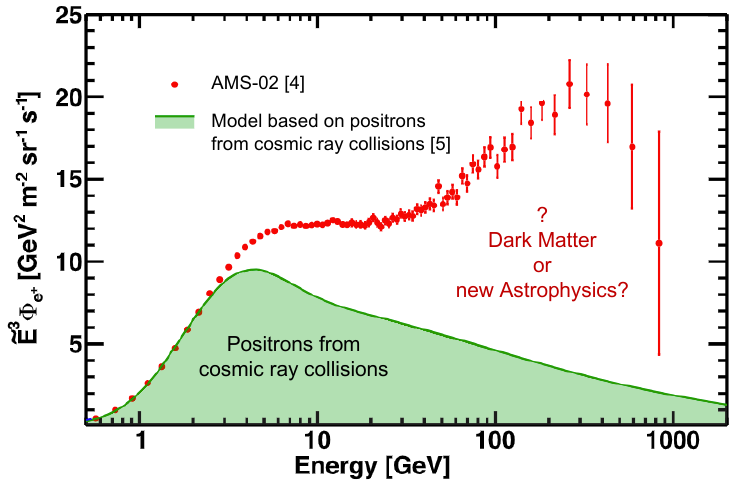}
         \caption{}
     \end{subfigure}
     \hfill
     \begin{subfigure}[b]{0.48\textwidth}
         \centering
         \includegraphics[width=\textwidth]{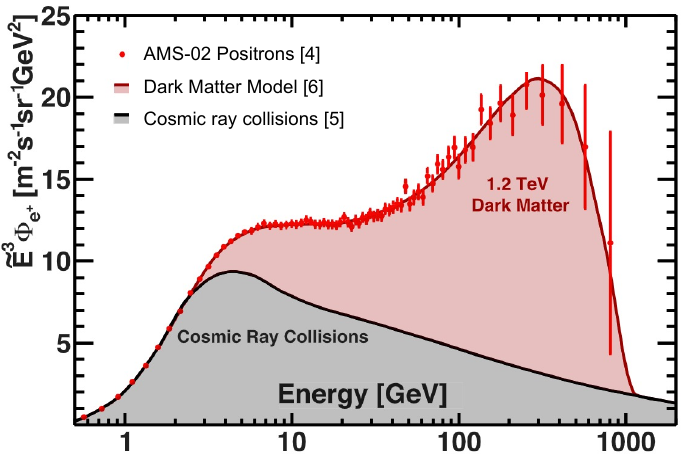}
         \caption{}
     \end{subfigure}
\caption[The cosmic positron flux measured by AMS.]{
The cosmic positron flux measured by AMS \cite{Aguilar2021TheYears} compared to (a) the theoretical results of the cosmic positron source being cosmic ray collisions \cite{Trotta2011ConstraintsAnalysis} and (b) a dark matter model \cite{Kopp2013ConstraintsResults}.
}
\label{fig:AMSPositrons}
\end{figure}

 One of the AMS's goals is to accurately measure the flux of cosmic positrons in space \cite{Aguilar2019TowardsPositrons}. Figure \ref{fig:AMSPositrons} shows the measurement of the cosmic positron flux detected by AMS \cite{Aguilar2021TheYears} and how it defers from the theoretical calculations of the flux from cosmic ray collisions \cite{Trotta2011ConstraintsAnalysis}, which is the main source of cosmic positrons. At high energies the results agree more closely with other theoretical models, for example positrons being produced from dark matter interaction \cite{Kopp2013ConstraintsResults}. As such, these results provide significant evidence for the existence of dark matter \cite{Aguilar2021TheYears}. 

\FloatBarrier

At present, the currently used analysis methods to separate cosmic positrons from the abundant proton background have a large uncertainty at energies close to 1 tera-electronvolts (TeV) and above. Of the two subdetectors capable of differentiating between positrons and protons, the TRD's proton rejection falls rapidly after 100 giga-electronvolts (GeV) \cite{Kirn2013TheStation} per charge of the particle (1 in the case of protons and positrons) and the ECAL can only partially contain electromagnetic showers at TeV energies \cite{Adloff2013TheCalorimeter}.

This paper proposes a Deep Learning (DL) approach to help improve positron classification for the ECAL. Using a multilayer perceptron (MLP), convolutional neural network (CNN), 2 residual neural networks (ResNets), and a Convolutional vision Transformer (CvT), we investigate the ability of deep learning to extract more information out of electromagnetic showers. To do so, we propose the treatment of the ECAL cells as pixels in a 3D array, in order to utilize the entire raw data output instead of handcrafted features, therefore treating cosmic ray shower as images. This allows the use of these state-of-the-art models to classify positrons and electrons, which present the same in ECALs, against protons. To our knowledge, this is the first paper that uses Convolutional vision Transformers as a means to classify signals in a particle physics detector. Additionally, we developed a physics-based feature engineering preprocessing method to improve the CvT's learning efficiency on calorimeter showers. Our code is will be released upon publication acceptance.

\section{Currently Used Methods}

The AMS currently uses two different models to differentiate between proton and positrons/electrons: the boosted decision tree (BDT) \cite{Krafczyk2016AStation} and the likelihood function (LHD) \cite{Kounine2017PrecisionCalorimeter}. Both models were trained on test beam data and used datasets that are no longer available for our use. As such, a fair comparison against our DL models is difficult to perform.

The BDT is a machine learning method that uses a regression based decision tree with the weights of misclassified events boosted \cite{Schapire2003TheOverview}. It is widely used and was often considered the de facto method for particle classification \cite{Roe2005BoostedIdentification} before the push for deep learning. AMS's BDT is trained on 19 correlated variables describing the ECAL shower such as shower maximum, width, energy in each layer and average location, and outputs a model score between -1 (protons) and +1 (positrons/electrons). The BDT achieves a proton rejection power of $10^3$ for particles with 1 TeV energy at 90\% electron efficiency \cite{Krafczyk2016AStation}.

The AMS LHD is a relatively newer method developed to better identify positrons at TeV energies when compared to the BDT model. It uses 7 main variables extracted from the shower (shower energy, x-, y-, and z-coordinate of the shower maximum, the zenith and azimuth angles that the shower makes, and the distance along the shower axis from the start of the shower to the shower maximum) along with 9 other correlated variables (number of cells in the shower, energy deposition around the shower axis for the first two layers, the energy deposition around the shower axis for the 3rd layer, etc.) to create an ECAL Likelihood, $\Lambda_{ECAL}$ that gives the log-likelihood of the shower being either a proton or a positron/electron. Tthe BDT achieves a proton rejection power of about $10^4$ for particles with 1 TeV energy at 90\% electron efficiency \cite{Kounine2017PrecisionCalorimeter}. A marked improvement over the AMS BDT.

\section{Methodology}

The ECAL output consists of energy depositions in 1296 cells organized into 18 layers with 72 cells in each layer. Every 2 layers, denoted as a superlayer, is dedicated to either the X or the Y direction. Four superlayers are dedicated to the X direction while five are dedicated to the Y direction \cite{Adloff2013TheCalorimeter}. This allows for 3D imaging of the ECAL shower and has a data format that resembles images.

We mainly use Monte Carlo (MC) simulations of cosmic ray protons and electrons/positrons to build datasets consisting of showers whose ECAL reconstructed energy ranges from 200 GeV to 2 TeV. As a final experiment, we train and test on actual data collected by AMS, denoted as ISS data, that have a reconstructed energy between 50 and 70 GeV.

The main deep learning architectures trained and tested are the MLP, CNN, ResNet \cite{He2016DeepRecognition}, and CvT \cite{Wu2021CvT:Transformers}. For MC, we first divide our dataset by separating the particles with a reconstructed energy below 1 TeV from the ones above 1 TeV. Then, we further divide each into train and test sets for below 1 TeV particles and validation (val) and test sets for above 1 TeV particles. We train the models on the train sets and validate after each epoch. This continues until 20 training epochs with no improvement of the model performance on the validation sets. If so, early stopping is employed and the trained model from the epoch with the best validation performance is saved. By training on below 1 TeV particles and validating on above 1 TeV particles, our goal was to train models that were less dependent on the total energy. By only training on lower energies, models that regularized, generalized and focused more on the shower shape would perform better on the higher energy events in the unseen validation set. This would allow for a higher confidence when the model classifies real positrons at high energies (TeV range) even without a large sample of labeled real data to test on. Finally, the saved models are tested on both the above and below 1 TeV test sets. For ISS data, the standard ML training, validation, and testing procedure was used. The datasets underwent a 60/20/20 split into train/validation/test sets, where the models were trained on the train set and evaluated on the validation set after each epoch.

\begin{figure}[!htbp]
    \centering
    \includegraphics[width=\textwidth]{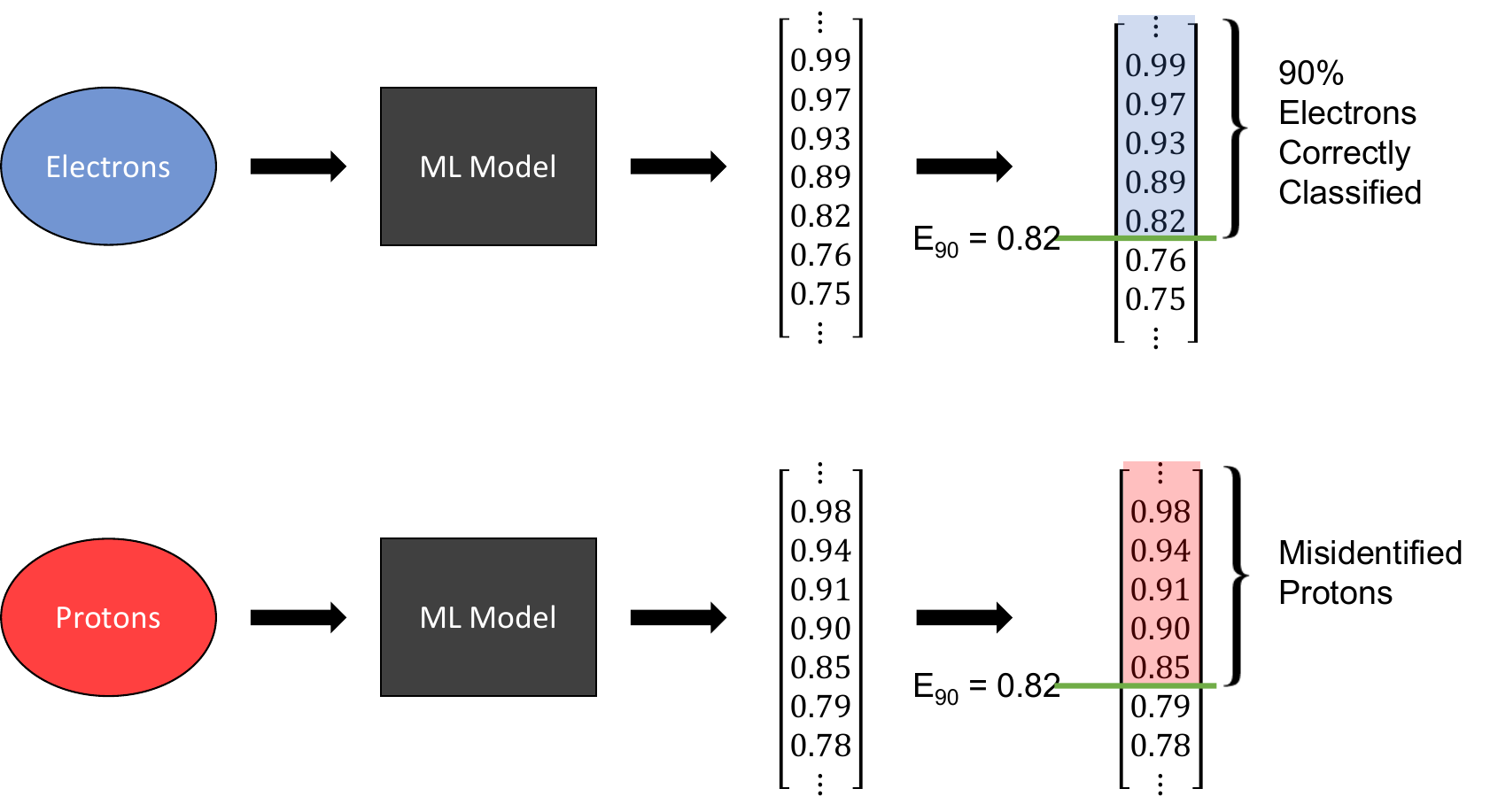}
    \caption[The process of how misidentified protons are calculated.]
    {
    The process of how misidentified protons are calculated. The dataset is first split into its constituent proton and electron datasets. The electrons are passed through the model being tested and the model's score for each electron event is recorded. Electron efficiency is the percent of electrons that are correctly classified given a specific model score cutoff, $E_{\%}$. Then, the protons are passed through the model and each proton's score is recorded. For a given $E_{\%}$, proton events that score above this value are classified incorrectly as electrons and are considered misidentified protons. 
    }
\label{fig:proton_rejection}
\end{figure}

The models are evaluated on the test set based on their proton rejection versus electron efficiency. Background rejection is a commonly used method to evaluate machine learning models for high energy physics experiments \cite{Chilingarian1994NeuralExperiments}\cite{Guest2016JetNetworks}\cite{Kekic2021DemonstrationExperiment}. For our work on the AMS ECAL, protons are considered the background \cite{Kounine2017PrecisionCalorimeter}. The process to find the misidentified protons for a dataset given a certain electron efficiency, $E_{\%}$ is shown in Figure \ref{fig:proton_rejection}. The dataset is first split into its constituent proton and electron datasets. The electrons are passed through the model being tested and the model's score for each electron event is recorded. Electron efficiency is the percent of electrons that are correctly classified given a specific model score cutoff, $E_{\%}$. Then, the protons are passed through the model and each proton's score is recorded. For a given $E_{\%}$, proton events that score above this value are classified incorrectly as electrons and are considered misidentified protons. The proton rejection for a given electron efficiency is then calculated as:

\begin{equation}
    \text{Proton Rejection} = \frac{\text{Total Number of Protons}}{\text{Number of Protons Misidentified}}
\end{equation}

This process is repeated for every electron efficiency from 1\% to 100\% and proton rejection versus electron efficiency figures are plotted. Additionally, proton rejection versus reconstructed energy is also plotted for the MC experiments. For a given percent of electron efficiency, often 60\% and 90\%, the proton rejection for each energy bin of the dataset is plotted to give an in-depth understanding of which energy bin the models perform poorly. 

\FloatBarrier

\section{Dataset}

\subsection{Monte Carlo}

\begin{figure}[!htbp]
     \centering
     \begin{subfigure}[b]{0.49\textwidth}
         \centering
         \includegraphics[width=\textwidth]{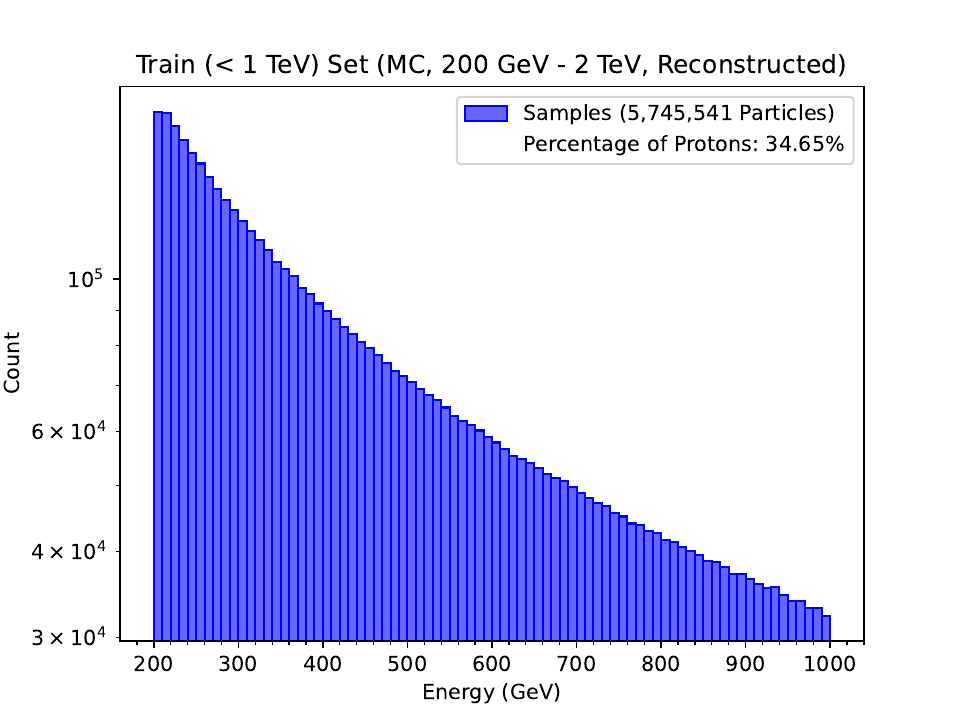}
     \end{subfigure}
     \hfill
     \begin{subfigure}[b]{0.49\textwidth}
        \centering
        \includegraphics[width=\textwidth]{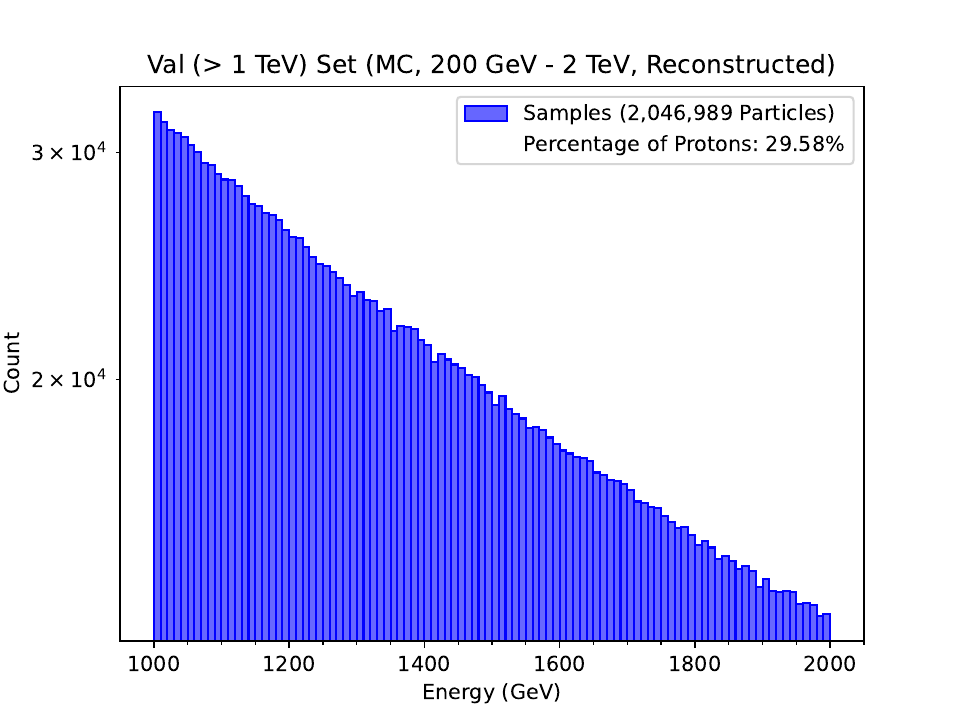}
     \end{subfigure}
     \hfill
     \begin{subfigure}[b]{0.49\textwidth}
         \centering
         \includegraphics[width=\textwidth]{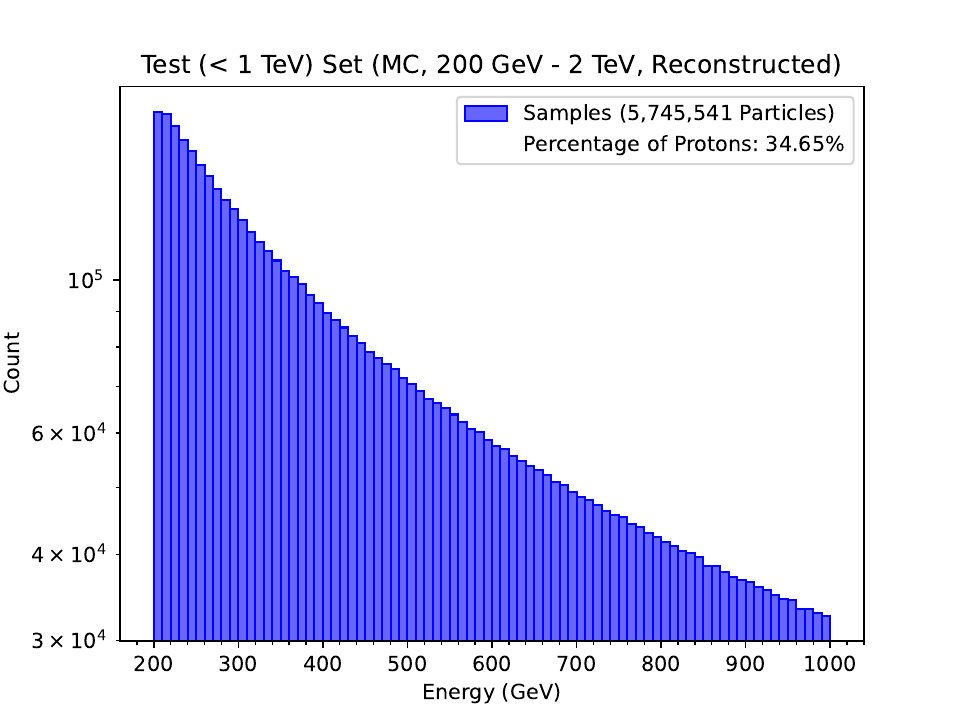}
     \end{subfigure}
     \hfill
     \begin{subfigure}[b]{0.49\textwidth}
         \centering
         \includegraphics[width=\textwidth]{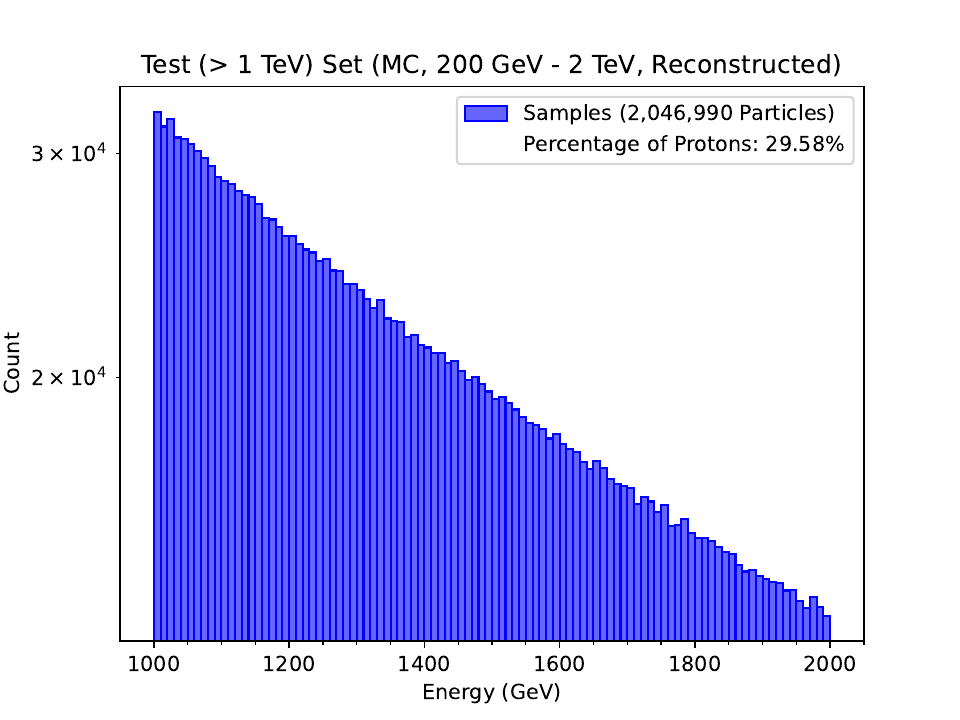}
     \end{subfigure}
\caption{Reconstructed energy histograms for the train, val, and two test sets, containing MC events with a reconstructed energy between 0.2–2 TeV.
}
\label{fig:Trk_2002000_stats}
\end{figure}

Our first dataset consists of MC events with a reconstructed energy between 200 GeV - 2 TeV. Figure \ref{fig:Trk_2002000_stats} shows the generated energy histograms of the train/val/test sets, to assure that energy distribution and class imbalance remains similar between the three sets. Since the events were generated using a power law similar to that observed in space, there are more lower energy events than higher energy events, making right-skewed histograms.

\FloatBarrier

\subsection{ISS Data}
\label{sec:ISS Data Extraction}

To extract ISS data, we used the TRD and Tracker to separate electrons from protons. The TRD in unable to accurately separate the two at high energies above 100 GeV \cite{Kirn2013TheStation}. In order to ensure a pure sample of electrons and protons with accurate labeling, we only extracted particles with a reconstructed energy between 50–70 GeV.

\begin{figure}[!htbp]
     \centering
     \begin{subfigure}[b]{0.49\textwidth}
         \centering
         \includegraphics[width=\textwidth]{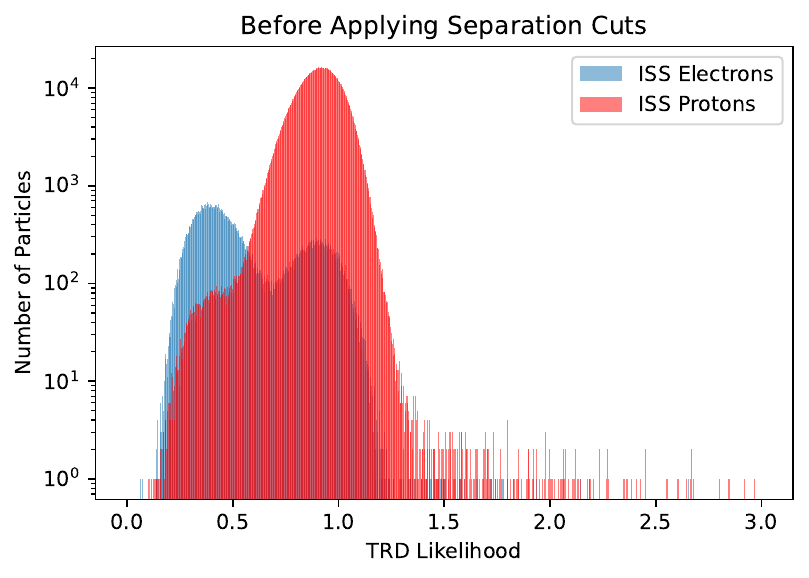}
         \caption{}
         \label{fig:BeforeCuts}
     \end{subfigure}
     \hfill
     \begin{subfigure}[b]{0.49\textwidth}
        \centering
        \includegraphics[width=\textwidth]{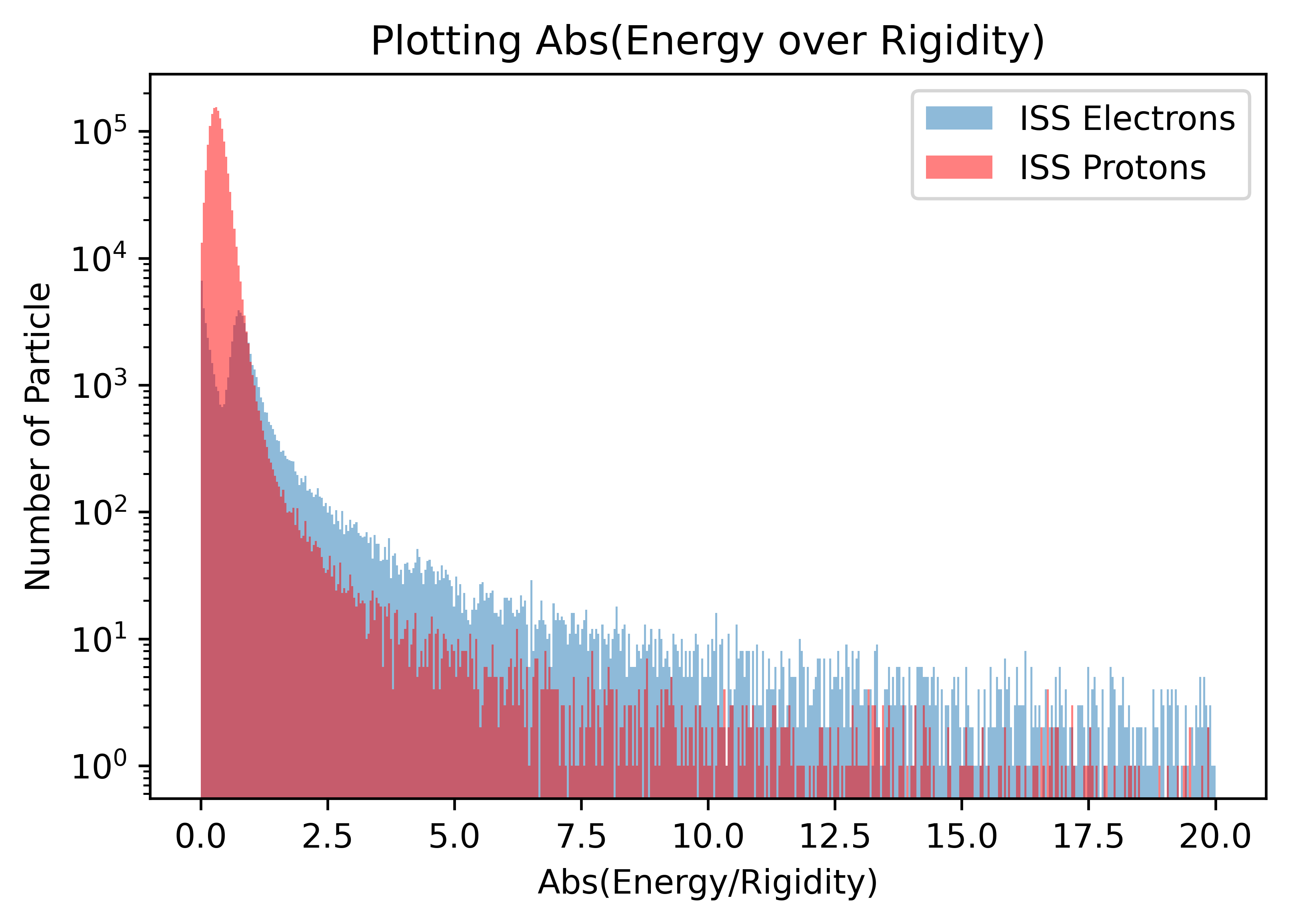}
         \caption{}
         \label{fig:Energy_Rig_Hist}
     \end{subfigure}
     \hfill
     \begin{subfigure}[b]{0.49\textwidth}
         \centering
         \includegraphics[width=\textwidth]{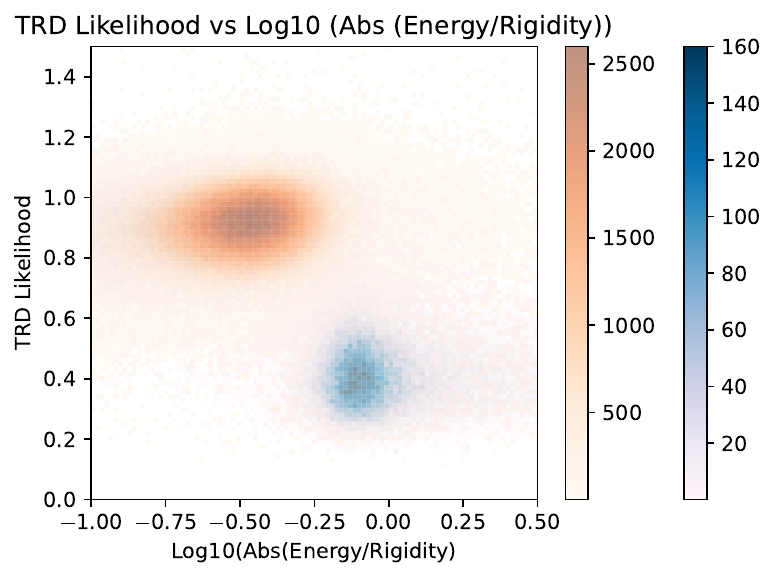}
         \caption{}
         \label{fig:LHD_Energy_Rig_Log10}
     \end{subfigure}
     \hfill
     \begin{subfigure}[b]{0.49\textwidth}
         \centering
         \includegraphics[width=\textwidth]{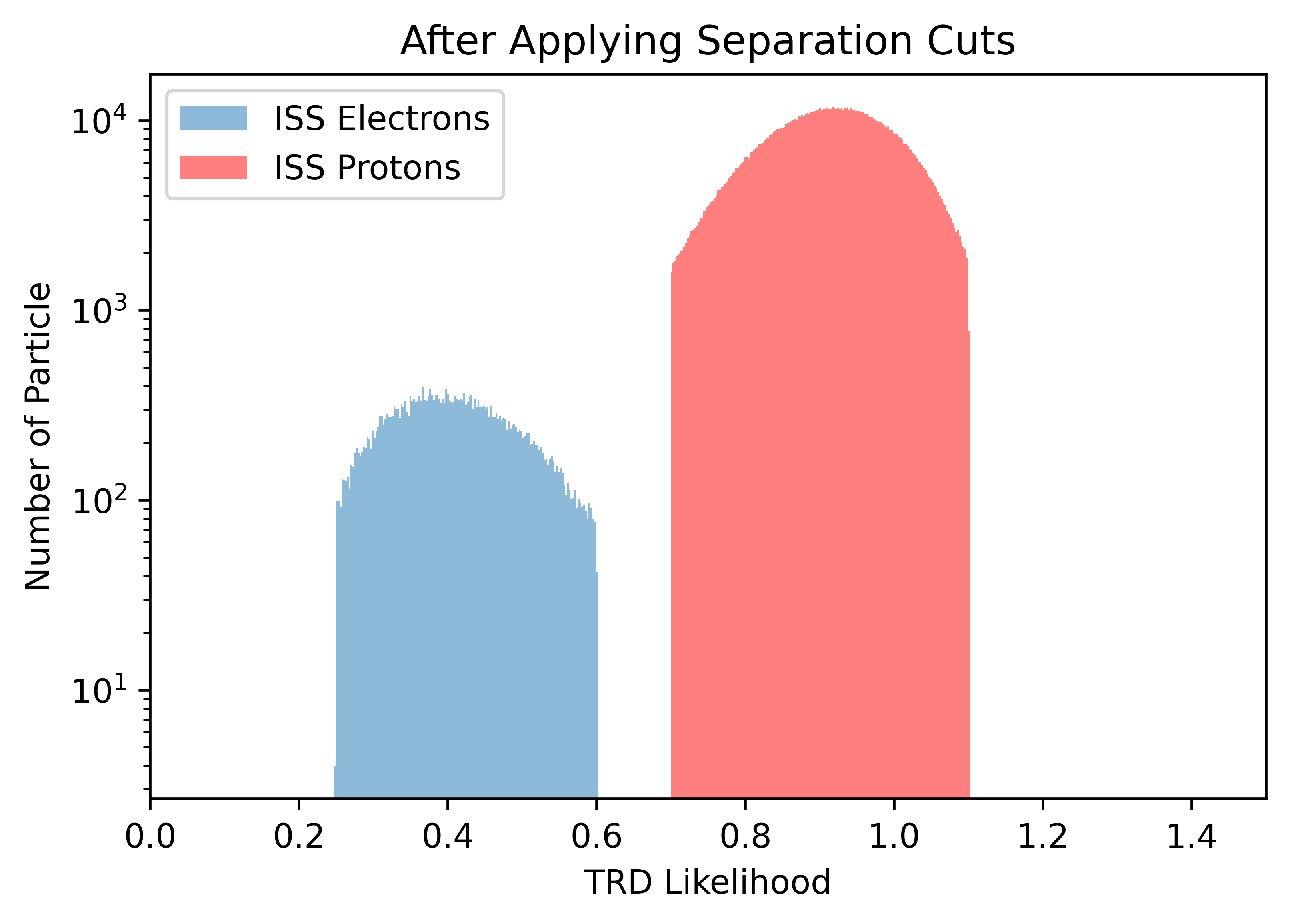}
         \caption{}
         \label{fig:AfterCuts}
     \end{subfigure}
\caption[Steps taken to separate ISS electrons from protons for the reconstructed energy range of 50–70 GeV.]{
Steps taken to separate ISS electrons from protons for the reconstructed energy range of 50–70 GeV. Due to the abundance of matter compared to antimatter, positively and negatively charged particles were labeled as ISS protons and electrons, respectively. The charge sign was determined using the Tracker. (a) Histogram of the TRD Likelihood. (b) Histogram of $|\frac{\text{Energy}}{\text{Rigidity}}|$ (c) 2D histogram of TRD Likelihood vs. $\log_{10} |\frac{\text{Energy}}{\text{Rigidity}}|$. (d) Histogram of the TRD Likelihood after learned cuts (filtering) on TRD Likelihood, energy, and rigidity is made.
}
\label{fig:ISS_Extraction}
\end{figure}

Figure \ref{fig:ISS_Extraction} depict the steps taken to separate electrons from protons. Due to the abundance of matter compared to antimatter, positively and negatively charged particles were considered and labeled as ISS protons and electrons, respectively. The TRD likelihood shows some separation of electrons from protons, as Figure \ref{fig:BeforeCuts} shows. Some overlap between positively and negatively charged particles are expected due to charge-confused protons \cite{AMSCollaboration2019ChargeAnalysis}, antiprotons, and positrons. The TRD likelihood is also highly dependent on energy and additional cuts on both energy, taken from the ECAL, and rigidity, taken from the Tracker, are required. The ECAL only accurately measures the energy of electrons and not protons, on account of it being designed for electromagnetic showers. The Tracker, however, can accurately measure the rigidity of both the electrons and protons well. As such, $|\frac{\text{ECAL Energy}}{\text{Tracker Rigidity}}| = 1$ for electrons, but should be lower for protons. 

Figures \ref{fig:Energy_Rig_Hist} shows a histogram of $|\frac{\text{ECAL Energy}}{\text{Tracker Rigidity}}|$ and Figure \ref{fig:LHD_Energy_Rig_Log10} shows a 2D histogram of the TRD Likelihood vs. $\log_{10} |\frac{\text{Energy}}{\text{Rigidity}}|$, where a clear separation of positively and negatively charged particles can be seen. Here, we note that electrons are indeed closer to $|\frac{\text{ECAL Energy}}{\text{Tracker Rigidity}}| = 1$.

\begin{figure}[!htbp]
     \centering
     \begin{subfigure}[b]{0.49\textwidth}
         \centering
         \includegraphics[width=\textwidth]{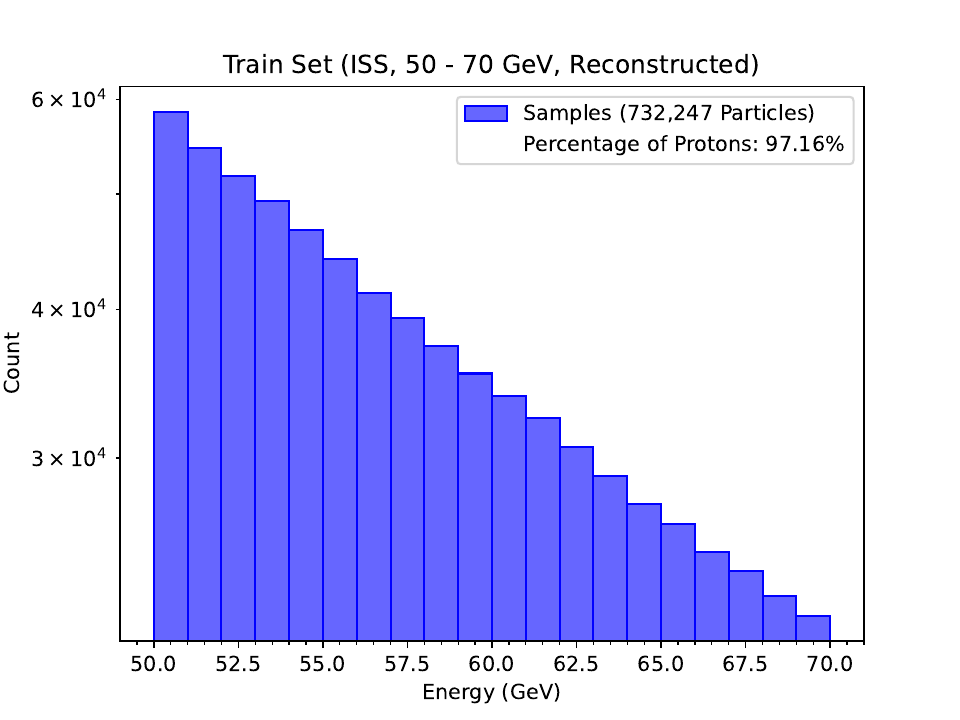}
     \end{subfigure}
     \hfill
     \begin{subfigure}[b]{0.49\textwidth}
        \centering
        \includegraphics[width=\textwidth]{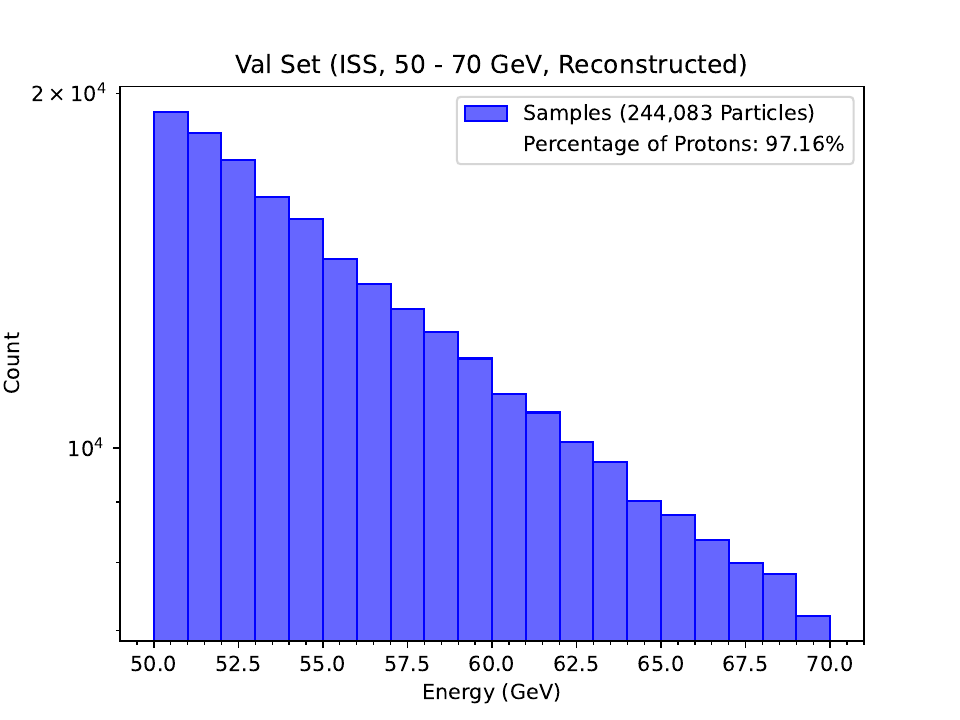}
     \end{subfigure}
     \hfill
     \begin{subfigure}[b]{0.49\textwidth}
         \centering
         \includegraphics[width=\textwidth]{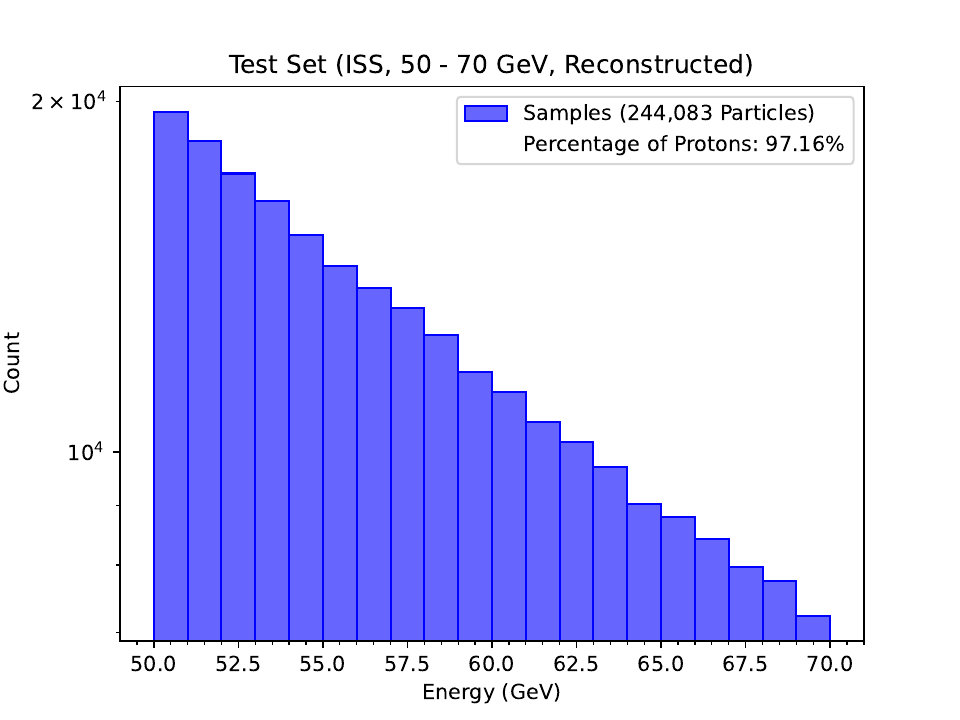}
     \end{subfigure}
\caption[Reconstructed energy histograms for the train/val/test sets created from the dataset consisting of ISS data with a reconstructed energy between 50–70 GeV.]{Reconstructed energy histograms for the train/val/test sets created from the dataset consisting of ISS data with a reconstructed energy between 50–70 GeV. The class imbalance (percentage of protons) and energy distribution is the similar between the three sets.}
\label{fig:ISS_stats}
\end{figure}

Figure \ref{fig:AfterCuts}, shows a selection of protons and electrons after the energy and rigidity cuts. Figure \ref{fig:ISS_stats} shows the energy histograms of the train/val/test sets created from ISS Data. The number of events in both the MC and ISS datasets can be seen in Table \ref{tab:dataset_stats}. We note the significantly smaller amount of data available in the ISS dataset, prompting our need to develop the physics-based feature engineering to improve learning efficiency for the CvT model. 

\FloatBarrier

\FloatBarrier

\begin{table}[]
\centering
\caption{Number of events (i.e. images) for each of the datasets.}
\label{tab:dataset_stats}
\resizebox{0.8\textwidth}{!}{%
\begin{tabular}{@{}llllll@{}}
\toprule
Source & Energy Range (GeV)                  & \multicolumn{2}{l}{Below 1 TeV (in Millions)} & \multicolumn{2}{l}{Above 1 TeV (in Millions)} \\ \midrule
       &                                             & Electrons             & Protons            & Electrons             & Protons            \\
MC     & 200-2000 & 7.51                  & 3.98               & 2.89                  & 1.21               \\
ISS    & 50-70                    & 0.03                  & 1.19               & 0                     & 0                  \\ \bottomrule
\end{tabular}%
}
\end{table}

\section{Models}

\begin{figure}[!htbp]
    \centering
    \includegraphics[width=0.35\textwidth]{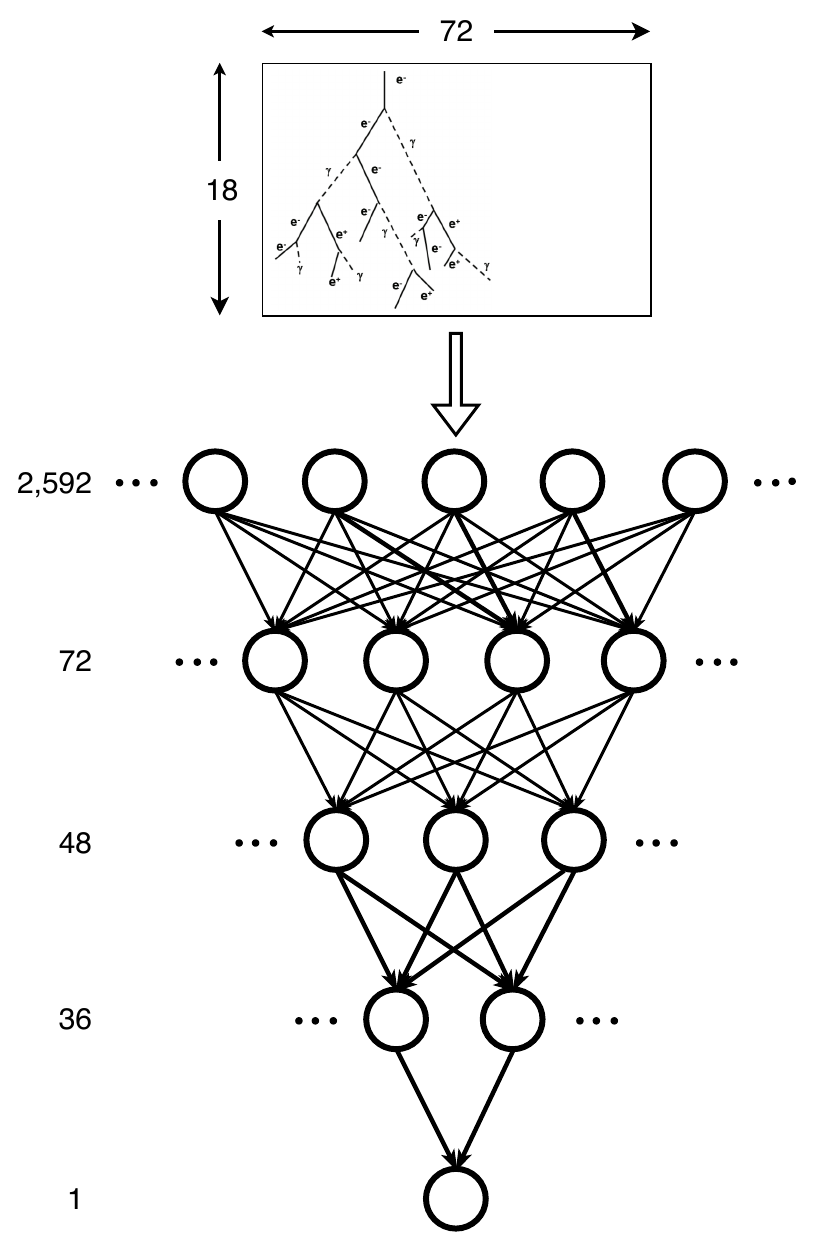}
    \caption{The architecture of our SimpleMLP model, with the number of neurons in each layer printed on the side.}
\label{fig:SimpleMLP}
\end{figure}

\begin{figure}[!htbp]
    \centering
    \includegraphics[width=0.7\textwidth]{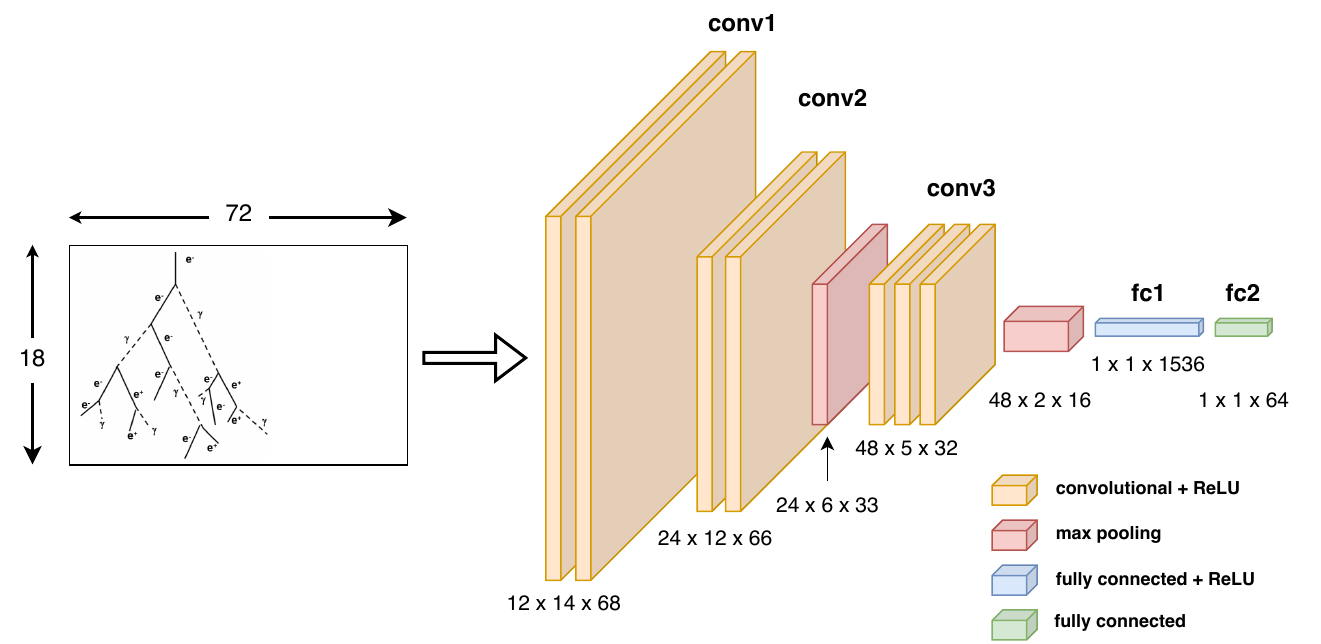}
    \caption{The architecture of our SimpleCNN Model. The dimensions of the resulting feature map are listed below each corresponding step.}
\label{fig:SimpleCNN}
\end{figure}

All of our models were designed to take in an $1\times18\times72$ input and to output a classification score. Our MLP, denoted as SimpleMLP, is shown in Figure \ref{fig:SimpleMLP} and consists of 4 hidden layers. Our CNN, denoted as SimpleCNN, is shown in Figure \ref{fig:SimpleCNN} and consists of 3 convolutions, 2 max pooling layers \cite{Ranzato2007UnsupervisedRecognition}, and 2 linear layers. For ResNets, we use two variants: the ResNet10 and ResNet18 \cite{He2016DeepRecognition}. For the ResNet18 variant, we used the default settings, modifiying only the initial and final layers. The ResNet10 model was custom built to create a model with half the number of trainable parameters, in order to make the number of parameters equivalent to the CvT's number of trainable parameters (see Table \ref{tab:model_capacity}).

Of particular interest is the CvT \cite{Wu2021CvT:Transformers}, a transformer model that takes in an input image and passes it to a convolutional layer (Convolutional Token Embedding) which reduces the resolution, increase the depth, and extracts feature vector tokens. The output is passed to a convolutional transformer block, where convolutions are used to build the Query (Q), Key (K), and Value (V) matrices that are then passed to the attention mechanism, where they are combined via the scaled dot-product attention,

\begin{align}
  \text{Attention}(Q, K, V) = \text{softmax}\left(\frac{QK^T}{\sqrt{d_k}}\right)V \label{eq:dot-product attention}
\end{align}

where $d_k$ is the dimension of K. More information on this can be found in the relevant literature \cite{Wu2021CvT:Transformers} \cite{Vaswani2017AttentionNeed}. The attention mechanism can learn to focus on regions of the feature map that are most discriminative for a class. The final output of the attention block is then passed through an MLP before being reshaped into a 2D feature map and passed to the next stage. At the final stage, a classification token (cls) is added, which is trainable feature vector that extract features most relevant to the classification task, before being passed to an MLP for classification.

The CvT combines the benefits of CNNs with that of Transformers, such as better generalization, better focus on key areas, and global context. This is important for ECAL showers as shower images can appear differently depending on the type of particle, energy of the particle, the angle of incidence and the point of entry.

We modified our implementation to better suit the AMS ECAL data by determining the number of transformer blocks that produces a consistent decrease in loss function compared to other configurations. Our final choice of configuration uses 4 transformer blocks, with Stage 3 having 2 blocks. For the three stages, there are 1, 3, and 6 attention heads, the kernel sizes chosen were $2 \times 6$, $3 \times 4$, and $3 \times 3$, and the stride lengths were $1 \times 2$, $1 \times 2$, and $1 \times 1$, respectively.

\FloatBarrier

\FloatBarrier


\subsection{Physics-based Feature Engineering}
\label{sec:feat_eng}

\begin{figure}[!htbp]
    \centering
    \includegraphics[width=\textwidth]{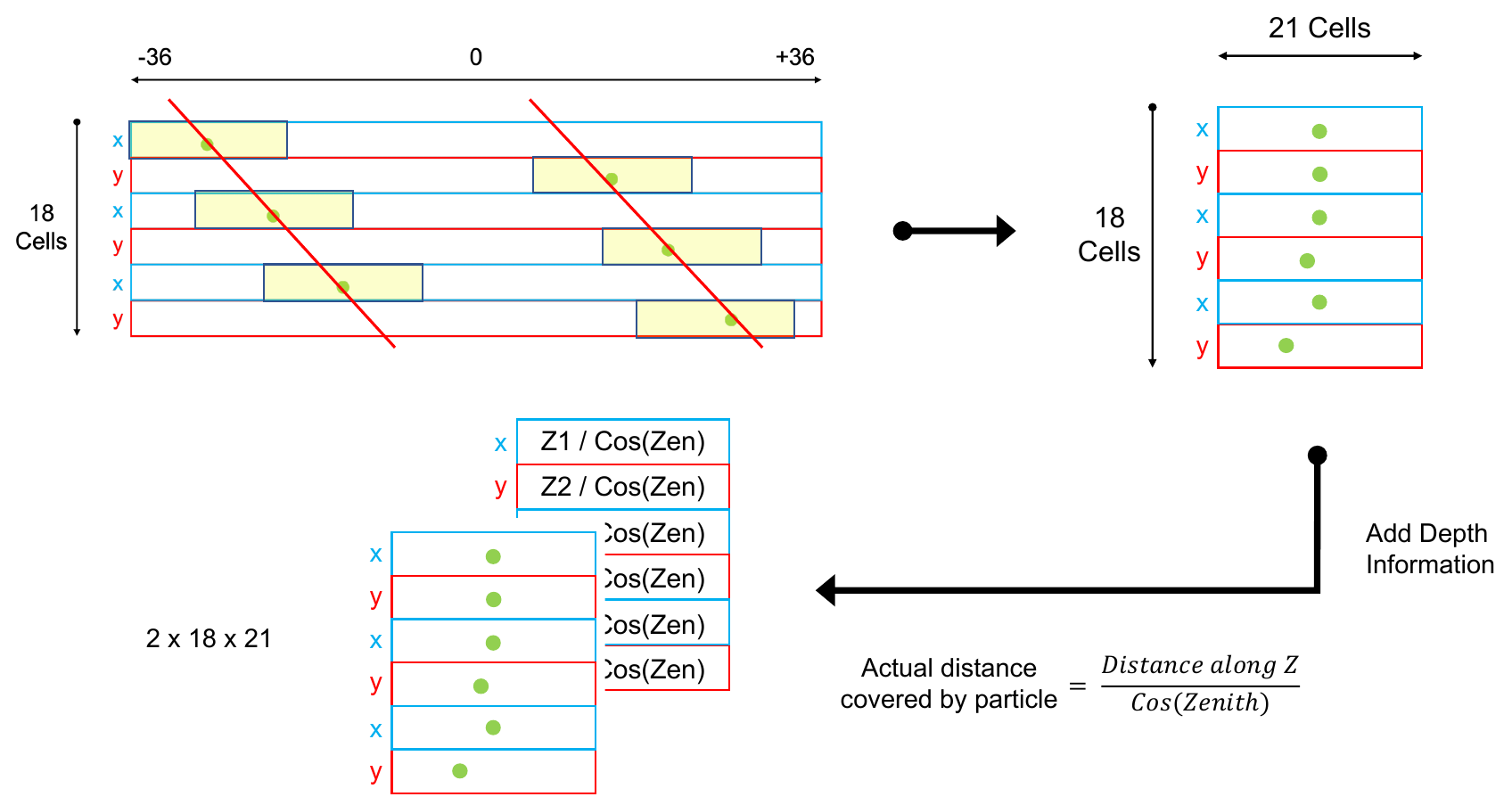}
    \caption[The physics-based feature engineering method.]
    {
    The physics-based feature engineering method. Using 5 variables from the Tracker (X, Y, Z, Zenith, and Azimuth of the particle incident on the ECAL), we project the path that the incident particle will follow (red lines). We then extract 10 pixels surrounding the projected locations in each layer (highlighed rectangles), create a 1 x 18 x 21 cell dataset, then add depth information as a second channel, resulting in a 2 x 18 x 72 shape. The depth information consists of the actual depth through the material crossed by the incident particle.
    }
\label{fig:Feature_Engineering}
\end{figure}

We developed a physics-based feature engineering method, shown in Figure \ref{fig:Feature_Engineering}, with the goal of making our CvT more efficient at learning on smaller training sets. This feature engineering uses the Tracker's projected line of the particle’s shower through the ECAL to provide an independent estimation of the shower direction. Each pixel along the projection, along with the 20 surrounding pixels, is extracted and made into another dataset of dimensions $1 \times 18 \times 21$. A second channel containing the actual depth crossed by the particle was also added. This was done by dividing the vertical distance of the $z$ layers by $\cos{\theta}$ to get the diagonal, true distance. Thus, the final dimension of the new feature engineered dataset is $2 \times 18 \times 21$. When using this in conjunction with the CvT, we label the model as Phys+CvT.

With all relevant models and variations being introduced, each model's total number of trainable parameters can be seen in Table \ref{tab:model_capacity}.

\FloatBarrier

\begin{table}[!htbp]
\centering
\caption{Total number of trainable parameters for each of the deep learning models.}
\label{tab:model_capacity}
\resizebox{0.4\textwidth}{!}{%
\begin{tabular}{@{}ll@{}}
\toprule
Model        & Trainable Parameters \\ \midrule
SimpleMLP    & 192,001              \\
SimpleCNN    & 106,317              \\
ResNet10 & 4,900,033            \\
ResNet18 & 11,170,753           \\
CvT      & 4,895,873            \\
Phys+CvT      & 4,896,641            \\ \bottomrule
\end{tabular}%
}
\end{table}

\subsection{Shared Configurations}
\label{sec:train_procedures}

Certain hyperparameters were shared between all the DL models for fairness. These can be seen in Table \ref{tab:shared_hyperparameters}. For each model, we searched for the optimal learning rate using grid search on a log-scale, and 1.0e-4 was found to be the best for all models. During training, early stopping was employed such that when patience exceeded the tolerance value, we would stop the training and save the trained model with the best score on the validation set. Two hardware setups were used interchangeably, shown in Table \ref{tab:hardware_config}. On average, the CvT and ResNet models would take 48 and 18 hours to train, respectively, before early stopping was triggered on these setups.

\begin{table}[!htbp]
\centering
\caption{Training hyperparameters for the deep learning models.}
\label{tab:shared_hyperparameters}
\resizebox{0.8\textwidth}{!}{%
\begin{tabular}{@{}ll@{}}
\toprule
Hyperparameter                       & Value                                     \\ \midrule
Batch Size                           & 128                                       \\
No. of Workers                       & 4                                         \\
Loss Function                        & Weighted Binary Cross Entropy with Logits \\
Activation Function after Last Layer & Sigmoid                                   \\
Optimizer                            & Adam                                      \\
Learning Rate for Optimizer          & 1.0e-4                                    \\
Early Stopping Patience              & 20                                        \\
Early Stopping Tolerance             & 1.0e-5                                    \\ \bottomrule
\end{tabular}%
}
\end{table}

\begin{table}[!htbp]
\centering
\caption{List of core components used to train the deep learning models. }
\label{tab:hardware_config}
\resizebox{0.8\textwidth}{!}{%
\begin{tabular}{@{}lll@{}}
\toprule
Name & IVMER DL                            & ImageLab DL                              \\ \midrule
GPU  & 2 x Nvidia RTX 2080Ti               & 2 x Nvidia RTX 2080Ti                    \\
CPU  & Intel Core i9-10980XE CPU @ 3.00GHz & Dual Intel Xeon CPU E5-2630 v3 @ 2.40GHz \\
RAM  & 256 GB DDR4                         & 189 GB DDR4                              \\ \bottomrule
\end{tabular}%
}
\end{table}

\section{Experiment and Results}
\subsection{Monte Carlo}

\begin{figure}[!htbp]
     \centering
     \begin{subfigure}[b]{0.49\textwidth}
         \centering
         \includegraphics[width=\textwidth]{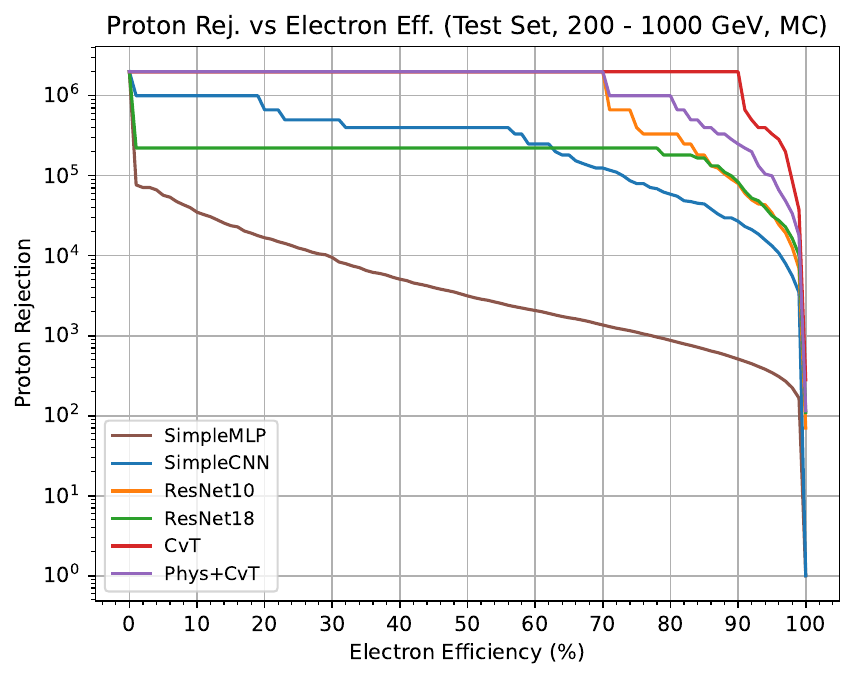}
         \caption{}
     \end{subfigure}
     \hfill
     \begin{subfigure}[b]{0.49\textwidth}
        \centering
        \includegraphics[width=\textwidth]{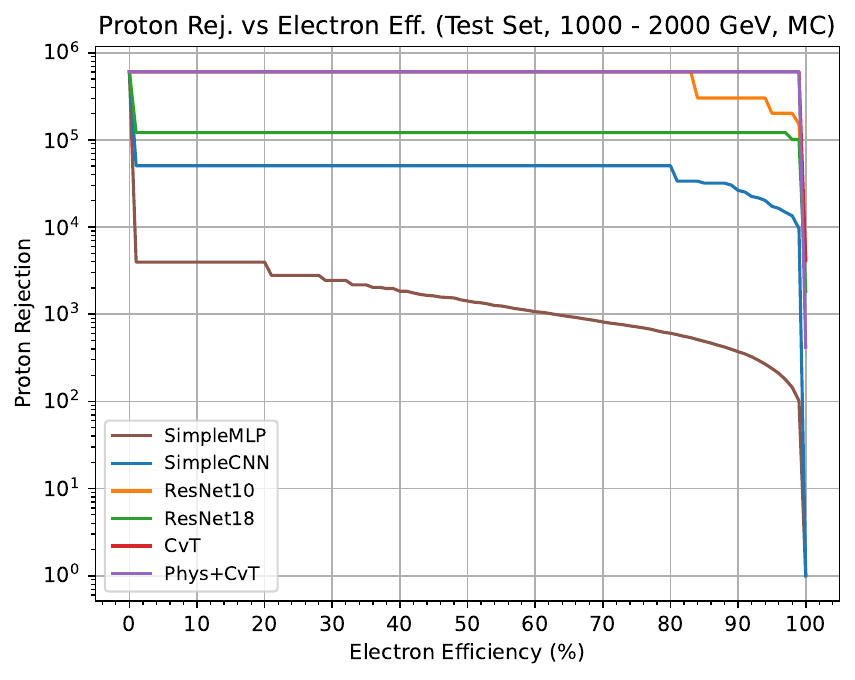}
         \caption{}
     \end{subfigure}
\caption[Proton rejection versus electron efficiency for the MC dataset.]{Proton rejection versus electron efficiency for the MC dataset. (a) At 90\% electron efficiency, below 1 TeV, the CvT outperforms the Phys+CvT, ResNet18, ResNet10, SimpleCNN, and SimpleMLP models by factors of 8, 24, 25, 74, and 3900, respectively. (b) At 90\% electron efficiency, above 1 TeV, the CvT performs equally with the Phys+CvT and outperforms the ResNet10, RestNet18, SimpleCNN, and SimpleMLP models by factors of 2, 5, 23, and 1630, respectively.}
\label{fig:Trk_2002000_pr_vs_eff}
\end{figure}

Figure \ref{fig:Trk_2002000_pr_vs_eff} shows the proton rejection versus electron efficiency plots. The CvT is the best performing model, with the Phys+CvT coming in second for below 1 TeV and practically equal for above 1 TeV. Additionally, below 1 TeV, ResNet10 outperforms the ResNet18 until 86\% electron efficiency, after which their performance becomes equal. Above 1 TeV, the ResNet10 outperforms all the non-CvT models at every electron efficiency.

\begin{figure}[!htbp]
     \centering
     \begin{subfigure}[b]{0.70\textwidth}
         \centering
         \includegraphics[width=\textwidth]{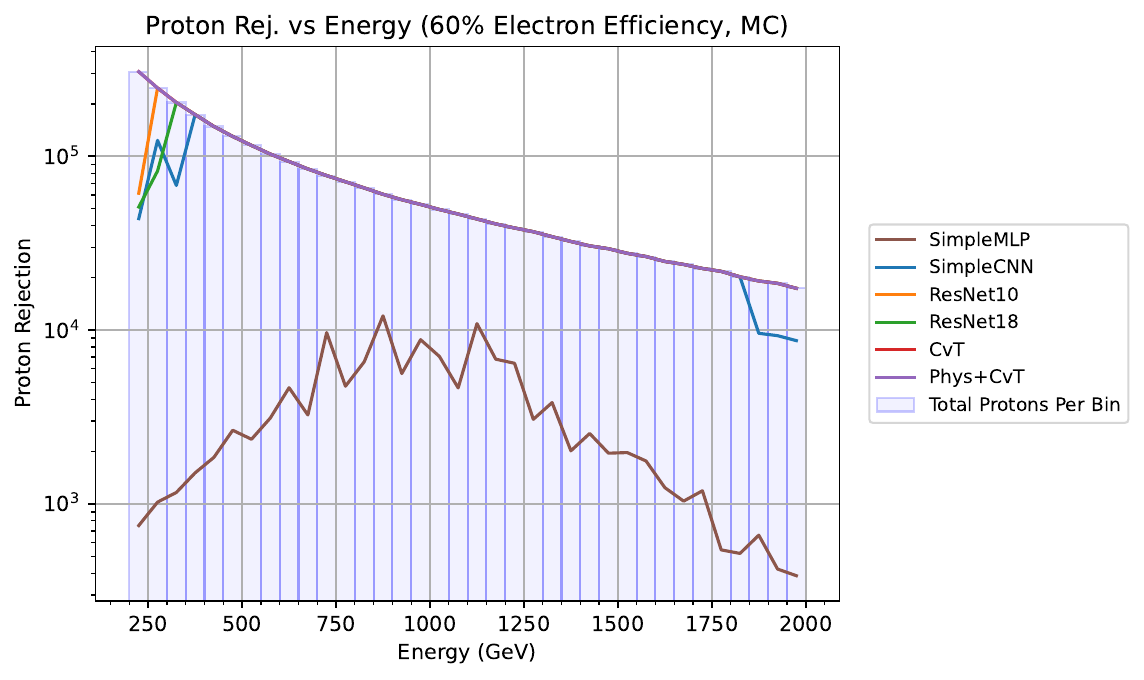}
         \caption{}
     \end{subfigure}
     \hfill
     \begin{subfigure}[b]{0.70\textwidth}
        \centering
        \includegraphics[width=\textwidth]{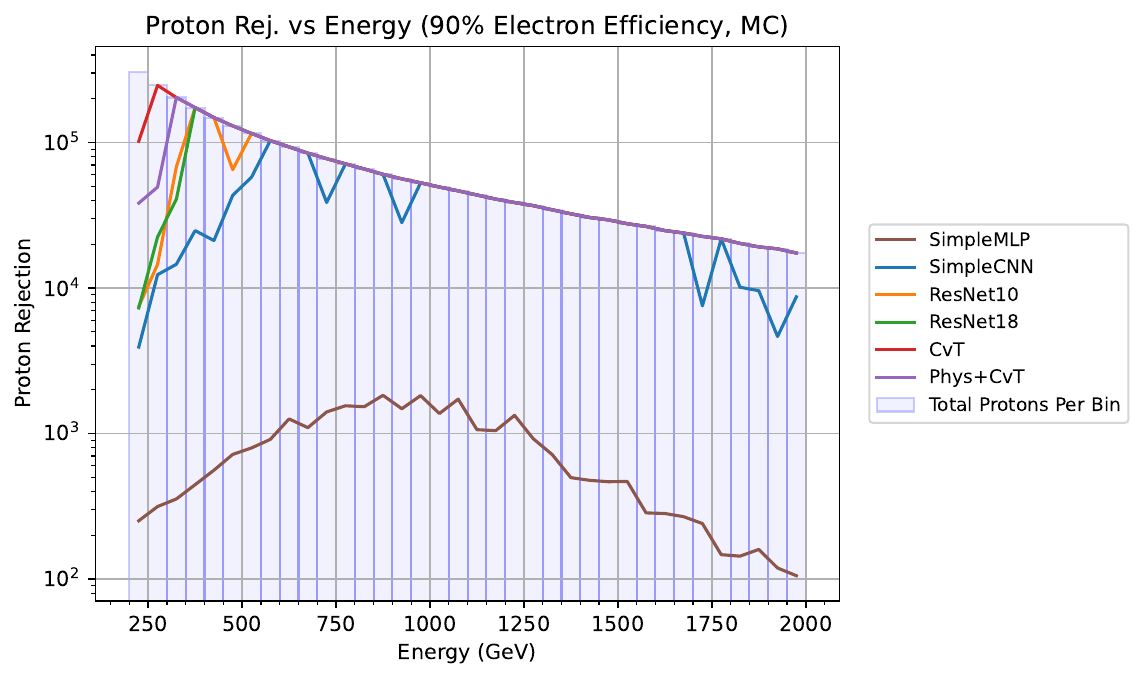}
         \caption{}
     \end{subfigure}
\caption[Proton rejection versus energy for the MC dataset.]{Proton rejection versus energy for the MC dataset. (a) At 60\% electron efficiency all the CvT models completely reject all the available protons. (b) At 90\% electron efficiency, the CvT performances drops for particles below 350 GeV, but are successful at rejecting all the protons above 350 GeV.}
\label{fig:Trk_2002000_pr_vs_energy}
\end{figure}

Figure \ref{fig:Trk_2002000_pr_vs_energy} shows the proton rejection for each energy bin at 60\% and 90\% electron efficiency. At 60\% electron efficiency, the CvT models reject all available protons. At 90\% electron efficiency, they only lose performance below 350 GeV. At 60\% electron efficiency, the ResNet18 starts to dip about 50 GeV before the ResNet10. At 90\% electron efficiency, while the ResNet10 starts to dip before ResNet18, the latter has a steeper decline, resulting in the performance seen in Figure \ref{fig:Trk_2002000_pr_vs_eff}.

The results show that the ResNet10 model boasts slightly better performance over the ResNet18, but does not match the performance of the CvT or Phys+CvT model. Between the latter two, the CvT outperforms the Phys+CvT below 1 TeV, but has equal performance for above 1 TeV. The SimpleMLP and SimpleCNN perform the worse out of all the tested models.

\FloatBarrier

\FloatBarrier

\subsection{ISS Data}



\begin{figure}[!htbp]
     \centering
     \begin{subfigure}[b]{0.46\textwidth}
         \centering
         \includegraphics[width=\textwidth]{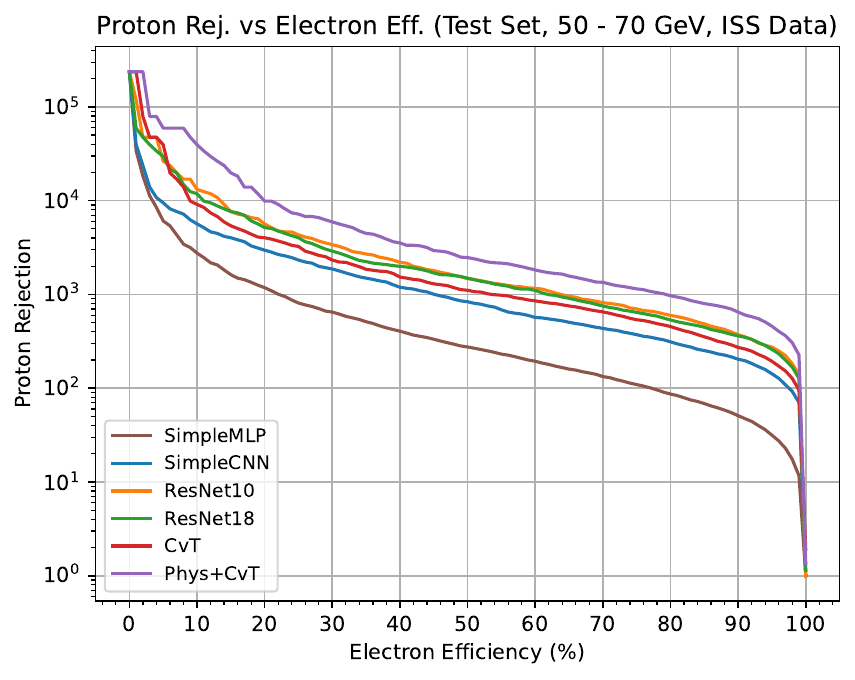}
     \end{subfigure}
     \hfill
     \begin{subfigure}[b]{0.50\textwidth}
        \centering
        \includegraphics[width=\textwidth]{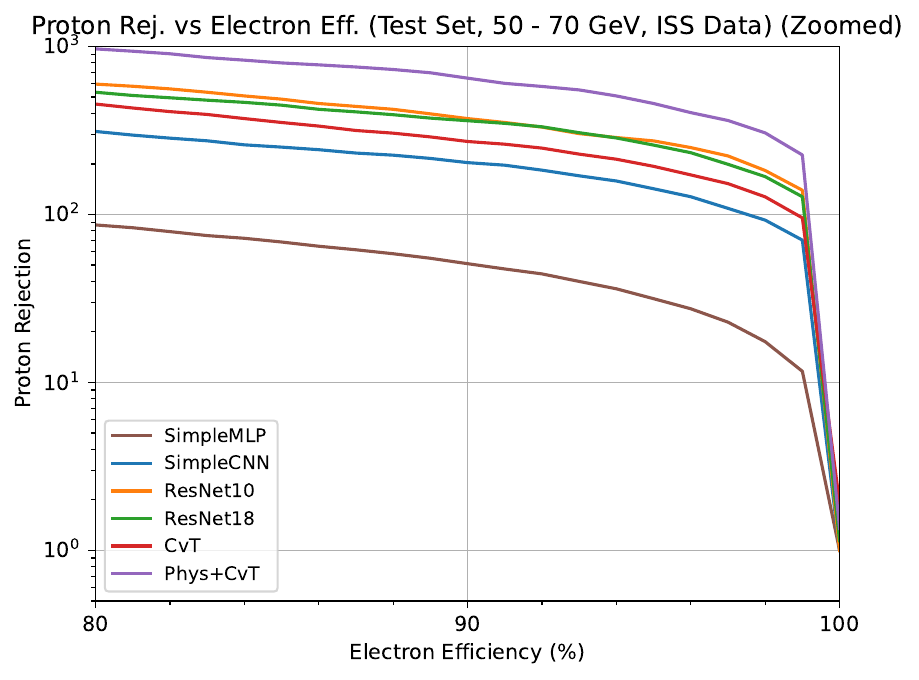}
     \end{subfigure}
\caption[Proton rejection versus Electron Efficiency for the ISS dataset.]{Proton rejection versus Electron Efficiency for the ISS dataset (left) and a zoomed in version for clarity (right). At 90\% electron efficiency, the Phys+CvT model outperforms the ResNet10, ResNet18, CvT, SimpleCNN, and SimpleMLP models by factors of 1.74, 1.79, 2.38, 3.18, and 12.72, respectively. 
}
\label{fig:ISS_Data_ISS_trained}
\end{figure}

We trained, validated, and tested our models on the ISS data. The proton rejection vs. electron efficiency plot in Figure \ref{fig:ISS_Data_ISS_trained} shows the Phys+CvT model outperforming the ResNet10, ResNet18, CvT, and SimpleCNN models by factors of 1.74, 1.79, 2.38, and 3.18, respectively. The results show that the Phys+CvT method was successful in being more data efficient on the ISS data. Due to the small energy range, the proton rejection vs. energy performance metric was not performed.


\FloatBarrier

\section{Conclusion}

From our experiment with the MC dataset, we saw the CvT models outperform all of the other deep learning models. This gives evidence of the transformer's ability to better generalize from below 1 TeV to above 1 TeV events, showing a reduced dependence on energy when compared to the other models. The Phys+CvT edition did not perform as well, indicating that with sufficiently available training data, the feature engineering was not very useful and indeed reduced performance on below 1 TeV events. Additionally, we note that the CvT outperformed the ResNet10 model even with a similar number of trainable parameters, indicating that overfitting was not the reason for the ResNet's relatively poor performance and providing evidence that the success of the CvT model was due to the architecture itself. 

From our experiment with the ISS dataset, we saw the Phys+CvT model outperforming the CvT model by a factor of 2.38. In fact, we note the better peformance of both the ResNet models over the CvT model, indicating the need for large amounts of training data in order for the CvT model to excel. As such, for this range of ISS data, the physics-based feature engineering improved learning performance for the CvT and that the Phys+CvT model shows potential to more accurately separate ISS positrons/electrons from protons.

In conclusion, we have provided empirical evidence of newer architectures, such as the CvT, being a viable alternative for particle classification in calorimeters, and that they show promise for future use in the AMS experiment.

\section*{Acknowledgements}
The authors would like to thank Professor Samuel C. C. Ting for his support, Dr. Zhili Weng for his invaluable guidance and comments, Berk Türk for his contributions in testing the CvT model, and both the AMS Collaboration and METU IVME-R their input and feedback. This research was supported in part by the Turkish Energy, Nuclear and Mineral Research Agency (TENMAK) under Grant No. 2020TAEK(CERN)A5.H1.F5-26. Some of the materials contained in this paper were previously published (in modified form) in a master's thesis \cite{Hashmani2023ACalorimeter}.

\printbibliography[title=REFERENCES]

\end{document}